\documentclass[twocolumn,aps,prl,superscriptaddress,amsmath,amssymb]{revtex4}
\usepackage{graphicx}

\begin{document}

\title{Evidence for the $S=9$ excited state in
Mn$_{12}$-bromoacetate measured by electron paramagnetic
resonance}

\author{K. Petukhov}
\affiliation{Department of Physics, University of Florida,
Gainesville, FL 32611--8440,USA}
\author{S. Hill}
\email[corresponding author, Email: ]{hill@phys.ufl.edu}
\affiliation{Department of Physics, University of Florida,
Gainesville, FL 32611--8440,USA}
\author{N. E. Chakov}
\affiliation{Department of Chemistry, University of Florida,
Gainesville, FL 32611--7200, USA}
\author{G. Christou}
\affiliation{Department of Chemistry, University of Florida,
Gainesville, FL 32611--7200, USA}
\author{K. Abboud}
\affiliation{Department of Chemistry, University of Florida,
Gainesville, FL 32611--7200, USA}

\date{\today}

\begin{abstract}
We present high-frequency high-field electron paramagnetic
resonance (EPR) measurements on the
[Mn$_{12}$O$_{12}$(O$_{2}$CCH$_{2}$Br)$_{16}$(H$_{2}$O)$_{4}$]
dodecanuclear manganese complex (Mn$_{12}$-BrAc). The
crystal-field parameters are found to be identical to those of the
original compound Mn$_{12}$-acetate
([Mn$_{12}$O$_{12}$(O$_{2}$CCH$_{3}$)$_{16}$(H$_{2}$O)$_{4}$]). A
detailed analysis of the frequency and temperature dependence of
anomalous peaks observed in the EPR spectra of Mn$_{12}$-BrAc
enables us to locate the $S=9$ manifold at about 40~K above the
$M_{S}= \pm 10$ ground state of this nominally $S=10$ system. This
is very close to the $M_S = \pm 6$ state of the $S = 10$ manifold,
thus suggesting pathways for the thermally assisted magnetization
dynamics and related properties. Finally, the EPR fine structures
recently attributed to disorder associated with the acetic acid of
crystallization in Mn$_{12}$-Ac are absent in the present
measurements, thus suggesting that the Mn$_{12}$-BrAc complex
represents a more suitable candidate for measurements of quantum
effects in high symmetry $S=10$ SMMs.
\end{abstract}


\maketitle

In 1996, it was reported that Mn$_{12}$-Ac exhibits resonant
magnetic quantum tunneling (MQT), as evidenced by steps that occur
at regular intervals in the low-temperature hysteresis loops of
the complex~\cite{Friedman96,Thomas96}. Since then, considerable
interest has been devoted towards the understanding of the quantum
behavior of single molecule magnets (SMMs), in particular
Mn$_{12}$-Ac (shorthand for Mn$_{12}$-acetate
[Mn$_{12}$O$_{12}$(O$_{2}$CCH$_{3}$)$_{16}$(H$_{2}$O)$_{4}$]$\cdot$2CH$_3$CO$_2$H$\cdot$4H$_2$O).
One of the most intriguing open questions concerns whether or not
one can ignore couplings to $S\neq10$ states in any theoretical
treatment of the low-temperature quantum properties of
Mn$_{12}$-Ac (including MQT). For example, recently the
low-temperature $S=9$ excited state was found in Fe$_8$Br$_8$,
another SMM with a $S=10$ ground state~\cite{Zipse}. In this paper
we provide clear experimental evidence for a $S=9$ excited state
of the nominally $S=10$ single-molecule magnet Mn$_{12}$-BrAc
(shorthand for Mn$_{12}$-\emph{bromo}acetate
[Mn$_{12}$O$_{12}$(O$_{2}$CCH$_{2}$Br)$_{16}$(H$_{2}$O)$_{4}$]$\cdot$4CH$_2$Cl$_2$),
a closely related complex to Mn$_{12}$-Ac.

Like Mn$_{12}$-Ac, which remains the most widely studied SMM
\cite{MRS00,Angewandte03}, Mn$_{12}$-BrAc crystallizes in a
tetragonal space group with individual molecules possessing $S_4$
site symmetry. Selection rules for quantum tunneling of the
magnetization imposed by the $S_4$ crystallographic symmetry are
not strictly obeyed by Mn$_{12}$-BrAc or Mn$_{12}$-Ac,
however~\cite{delBarcoPRL}. A number of explanations have been
proposed to account for this anomalous behavior, including crystal
dislocations~\cite{Garanin} and, more recently, disordered solvent
molecules of crystallization that give rise to symmetry breaking
effects~\cite{delBarcoPRL, HillPRL03a, HillCM, Cornia}. Cornia
\textit{et al.} have proposed that the fourfold ($S_4$) molecular
symmetry is disrupted by a strong hydrogen-bonding interaction
between an acetate ligand of the Mn$_{12}$ cluster and a
disordered acetic acid molecule of crystallization~\cite{Cornia}.
Using their model, up to six isomers can exist in the lattice,
each of which differs in the number and arrangement of
hydrogen-bonding interactions. Of the six isomers, only two have
crystallographically four-fold ($S_4$) symmetry. The remaining
four isomers have lower symmetry and can explain many experimental
factors associated with the low temperature hysteresis loops of
the complex, including odd-to-even M$_S$ MQT steps. It is
therefore of importance to study the magnetic behavior of a high
symmetry Mn$_{12}$ cluster that consist of only one species with
strict axial symmetry. The only other known complex to date that
meets these specifications is Mn$_{12}$-BrAc. The core of the
molecule is the same as that of Mn$_{12}$-Ac and the molecule also
possesses a spin $S = 10$ ground state. The complex is therefore
an ideal candidate for study, as there are four relatively inert
CH$_{2}$Cl$_{2}$ (dichloromethane) solvent molecules per
Mn$_{12}$-BrAc molecule. This is in contrast to the strongly
hydrogen-bonding nature of the water and acetic acid molecules of
crystallization in the case of Mn$_{12}$-Ac. However, like all
other known Mn$_{12}$-based SMMs, all tunneling resonances are
observed, including odd-to-even M$_S$ MQT steps \cite{delBarco}.
Recently, del Barco \textit{et al.} proposed that a distribution
of internal transverse magnetic fields in Mn$_{12}$-BrAc is
responsible for a lack of any selection rules in the MQT
phenomena~\cite{delBarco}.


The Mn$_{12}$-BrAc molecule consists of four Mn$^{4+}$ ions, each
with spin $S = \frac{3}{2}$, surrounded by eight Mn$^{3+}$ ions
with spin $S = 2$~\cite{MRS00,Angewandte03}. The orbital moment is
quenched, and a Jahn-Teller distortion associated with the
Mn$^{3+}$ ions is largely responsible for the magnetic anisotropy.
A simplified treatment of the magnetic interactions within the
molecule has been developed~\cite{SessoliJACS93,Katsnelson99,
RaghuPRB01, YamamotoRPL02, RegnaultPRB02} wherein four strongly
antiferromagnetically coupled Mn$^{3+}-$Mn$^{4+}$ dimers, each
with spin $S = 2 - \frac{3}{2} = \frac{1}{2}$, couple via an
effective ferromagnetic interaction to the four remaining $S = 2$
Mn$^{3+}$ ions, giving a total spin $S = 10$. The magnetic energy
levels of the rigid $S=10$ spin system are then usually described
by the effective single-spin Hamiltonian
\cite{Barra97,MirebeauPRL99}:


\noindent{ \hfill  \hfill  \hfill  \hfill  \hfill $\hat H = D\hat
S_z^2 + \mu _B \vec B \cdot
\mathord{\buildrel{\lower3pt\hbox{$\scriptscriptstyle\leftrightarrow$}}
\over g}  \cdot \hat S + \hat H', \hfill\hfill\hfill\hfill\hfill
(1)$}


\noindent{where $D$ $(< 0)$ is the uniaxial anisotropy constant,
the second term represents the Zeeman interaction with an applied
field $\vec{B}$
($\mathord{\buildrel{\lower3pt\hbox{$\scriptscriptstyle\leftrightarrow$}}
\over g}$ is the Land$\acute{e}$ g tensor), and $\hat{H}^\prime$
includes higher order terms in the crystal field ($\hat{O}^0_4$,
$\hat{O}^2_2$, $\hat{O}^2_4$, $\hat{O}^4_4$, {\em etc.}), as well
as environmental couplings such as intermolecular dipolar and
exchange interactions~\cite{HillPRB02,KPark02b}. The leading term
in Eq.~1 is responsible for the energy barrier to magnetization
reversal and the resulting magnetic bi-stability
\cite{MRS00,Angewandte03}. The weaker couplings between the four
spin$-\frac{1}{2}$ dimers and the four spin$-2$ Mn$^{3+}$ ions
largely determine the low energy excitations within the molecule
(to $S\neq 10$ states)~\cite{Katsnelson99,YamamotoRPL02}. However,
the nature of these couplings is not well known.

The inadequacy of the $S = 10$ model was perhaps first raised by
Caneschi {\em et al.} in  order to interpret the temperature
dependent susceptibility of Mn$_{12}$-Ac~\cite{Caneschi92}. Their
calculation proposed the existence of two degenerate $S=9$ states
at 0.725~THz ($\sim 35$~K), one $S=8$ state at 1.195~THz ($\sim
57$~K), and other $S \leq 8$ and $S > 10$ states at higher
energies. Not all of these states, however, were clearly observed
in the inelastic neutron scattering study of Mn$_{12}$-Ac, where a
well-pronounced mode was found around 1.2~THz, while the mode at
0.725~THz was hardly visible and the authors attributed this fact
to a very small intensity at the corresponding scattering vector
\cite{Hennion97}. Magnetization measurements have shown the
existence of $S \leq 9$ excited levels at energies between 30 and
90~K~\cite{BarbaraMMM98}. Furthermore, measurements of the
Mn$_{12}$-Ac transmission spectra in the submillimeter range
indicate the existence of some weak band around 30-35~cm$^{-1}$
($\sim 43$-50~K) at high temperatures, whose frequencies do not
match the transitions within the ground $S=10$ multiplet
\cite{MukhinEPL98}. Finally, by means of $^2$D and $^{13}$C NMR
investigations~\cite{AcheyPRB01,AcheySSC02} and $^{55}$Mn
spin-lattice relaxation measurements~\cite{YamamotoRPL02} it has
also been clearly established that the unpaired electron density
is distributed over the entire Mn$_{12}$-Ac framework. These
challenging experimental findings stimulated subsequent
calculations, which placed the $S = 9$ manifold at $35$~K or
higher~\cite{Katsnelson99,RegnaultPRB02,Park03}. All of these
observations provided the impetus for the present undertaking $-$
a detailed investigation of the spin-energy levels of a Mn$_{12}$
cluster by variable frequency, variable temperature EPR
spectroscopy. We provide definitive evidence for the existence of
a state at $40\pm 2$~K, with all of the properties of an $S = 9$
manifold. The results help the understanding of many of the
above-mentioned features and are in good agreement with more
recent high precision electronic structure calculations
\cite{Park03}.

Multi-high-frequency (51.5, 65.4, and 76.9~GHz) single crystal EPR
measurements were carried out using a millimeter-wave vector
network analyzer (MVNA) and a high sensitivity cavity perturbation
technique; this instrumentation is described in detail elsewhere
\cite{MolaRSI}. The measurements were performed using commercial
superconducting solenoid capable of producing fields up to 9~T. In
all cases, the temperature was stabilized ($\pm 0.01$~K) relative
to a calibrated Cernox$^{\mathrm{TM}}$ resistance sensor. We have
elaborated our cavities with rotatable endplates for sample
orientation with 0.18$^\circ$ accuracy~\cite{Takahashi}.
Submillimeter-sized single Mn$_{12}$-BrAc crystals were prepared
from Mn$_{12}$O$_{12}$(O$_{2}$CMe)$_{16}$(H$_{2}$O)$_{4}$
(Mn$_{12}$-Acetate) by a ligand substitution procedure involving
the treatment of Mn$_{12}$-Ac with an excess of
BrCH$_{2}$CO$_{2}$H in two cycles. Repeated treatment is necessary
as the ligand substitution reaction is an equilibrium process that
must be driven to completion if the pure product is to be obtained
\cite{Soler01, An, Tsai}. The initial complex Mn$_{12}$-Ac was
synthesized using standard methods~\cite{Lis80}. In order to
confirm the solvent content of 4$\cdot$CH$_{2}$Cl$_{2}$ per
formula unit, an X-ray structural refinement analysis was carried
out on a wet single-crystals removed directly from the mother
liquor. In order to avoid solvent loss in the EPR measurements,
the sample was removed from the mother liquor, immediately sealed
in silicone grease, and quickly transferred to the cryostat (ca. 5
minutes) where it was cooled under atmospheric pressure helium
gas. Similar sample handling procedures have been carried out for
Mn$_{12}$ samples containing considerably more volatile solvent
molecules, and these have shown minimal solvent loss, as verified
by before and after AC SQUID measurements \cite{Soler02}.

In order to align the sample's hard plane ($x,y$) with respect the
applied DC magnetic field, angle dependent measurements of the EPR
spectra were performed, as shown in Fig.~1(a). In general, the
hard-plane spectra of Mn$_{12}$-BrAc look very similar to those of
Mn$_{12}$-Ac \cite{HillPRL98,Edwards}. In the high-frequency limit
($g \mu_B \mathbf{B} >|D|S $), one expects a total of 20 EPR
transitions within the $2S + 1$ $(S = 10)$ multiplet, as shown in
the Fig.~1(b) by solid curves. The $\alpha$-resonances, which
correspond to transitions within the Zeeman-split M$_S = \pm m$
$(m =$~integer, and $0 \leq m \leq S$) zero-field levels, comprise
half of this total; in the zero-field limit, the quantization axis
is defined by the uniaxial crystal field tensor, and is along $z$.
In the high-field limit, the quantization axis points along the
applied field vector; the 10 $\alpha$-resonances then correspond
to transitions from M$_S =$~even-to-odd $m$, {\em e.g}. M$_S =
-10$ to $-9$. Because the $\alpha$-resonances originate from pairs
of levels ($\pm m$) which are (approximately) degenerate in zero
field, one expects the resonance frequencies, when plotted against
field, to tend to zero as the field tends to zero, as can be seen
in the Fig.~1(c). The simulation depicted in the Fig.~1(c) has
been performed by exact diagonalization of Eq.~1, and this
procedure is described in detail elsewhere~\cite{Edwards,
HillPRL98}. In order to fit our experimental data (open circles),
we have used crystal field (CF) parameters obtained from our
earlier studies of Mn$_{12}$-BrAc with the field along the easy
axis~\cite{EdwardsUNPB}: $D=-0.456$~cm$^{-1}$, $B^0_4 =-2.0 \times
10^{-5}$~cm$^{-1}$. These Hamiltonian parameters are very close to
the accepted CF parameters for Mn$_{12}$-Ac ($D=-0.454$~cm$^{-1}$,
$B^0_4 =-2.0 \times 10^{-5}$~cm$^{-1}$)~\cite{MirebeauPRL99,
Edwards, HillPRL03a}, thus emphasizing the close similarity of
physical properties of these two derivatives of the
Mn$_{12}$O$_{12}$ molecule.

In earlier investigations of Mn$_{12}$-Ac~\cite{HillPRL98,Edwards}
it was pointed out that EPR spectra obtained for a field applied
perpendicular to the easy axis of the molecule revealed a number
of anomalous transitions which were labeled $\beta$~\cite{note1},
as opposed to the $\alpha$-resonances which nicely fit the
accepted $S=10$ Hamiltonian (Eq.~1). Initially, these
$\beta$-transitions were tentatively ascribed to the
M$_S=$odd-to-even transitions (e.g. {\em e.g}. M$_S = -9$ to $-8$)
\cite{HillPRL98}; however we note here that they should not be
observable below a cut-off frequency, which is about 95~GHz at
high fields for the given CF parameters, as depicted in Fig.~1(c).
In full agreement with these calculations we \emph{do not} observe
these $\beta$-resonances until we slightly misalign the sample's
hard plane with respect to the applied magnetic field (Fig.~1(a)).
Indeed, at $\pm3.6^\circ$ away from the hard plane, the
$\beta$-resonances become highly pronounced. Meanwhile, the
$\alpha10$, $\alpha8$ and $\alpha6$ resonances disappear over this
same angle range, as shown in Fig.~1(a). In fact, there is an
approximately $0.75^\circ$ range over which neither $\alpha10$ or
$\beta9$ are observed and, although $\alpha4$ and $\alpha2$ peaks
remain visible at $\theta = \pm 3.6^\circ$, it is clear that their
intensities diminish substantially. This symmetry effect between
the out-of-plane angle dependence of the $\beta$ and $\alpha$
resonances was recently reported for Mn$_{12}$-Ac, and is
discussed in more detail in Ref.~\cite{HillCM}. For comparison,
Fig.~1(b) shows simulations of the EPR spectra for the same angle
range, generated using the program SIM~\cite{Weihe}. These
simulations agree more-or-less perfectly with our observations,
i.e. the $\alpha$ peaks disappear, and $\beta9$ appears, as the
field is tilted away from the hard plane. Indeed, the simulations
predict quite accurately the angles at which the $\alpha$ peaks
switch off, and $\beta9$ switches on. This contrasts the behavior
seen in Mn$_{12}$-Ac, where a significant overlap of the $\alpha$
and $\beta$ peaks has been attributed to a distribution of tilts
of the easy axes of the molecules (up to $\pm 1.7^\circ$), induced
by a discrete disorder associated with the two acetic acids of
crystallization. Although a small distribution of tilts can be
inferred from the present data, due to the absence of some of the
features in the simulations which disperse strongly with angle,
the width of the distribution must be small (of order the angle
step in these measurements, i.e. $0.2^\circ$). This indicates that
the distributions of transverse fields recently reported by
del~Barco \textit{et al.}, must have an explanation unrelated to
easy axis tilting. It is also apparent from the data in Fig.~1(a)
that, for the most part, the resonances are extremely symmetric
and much sharper than those observed for Mn$_{12}$-Ac (see
Ref.~\cite{HillCM}). In particular, none of the fine structures
seen in the hard axis spectra of Mn$_{12}$-Ac are seen for the
present complex, e.g. the pronounced high-field shoulders on the
$\alpha$ resonances. Consequently, we can conclude that the
discrete solvent disorder that is now well established in
Mn$_{12}$-Ac~\cite{Cornia,HillPRL03a, delBarcoPRL,HillCM} is
absent in Mn$_{12}$-BrAc. This observation is consistent with the
full compliment of four solvent molecules per formula unit, and
suggests that the BrAc complex probably represents a more suitable
candidate for measurements of quantum effects in high symmetry
$S=10$ SMMs.

In Fig.~2 (bottom panel) we present the temperature dependence of
the EPR spectra of Mn$_{12}$-BrAc at 51.5~GHz for a field applied
\emph{exactly} perpendicular to the easy axis of the molecule
($\mathbf{B} \perp z$). As the temperature is increased from 10~K
up to 40~K, an extra resonance is found at 7.42~T. Since this peak
is located between $\alpha10$ and $\alpha8$ resonances we have
labeled this resonance $\alpha9$. We have also observed similar
extra peaks at 65.4 and 76.9~GHz, which are located respectively
at 7.77 and 8.22~T between the corresponding $\alpha8$ and
$\alpha10$ resonances (although, according to the calculations
depicted in Fig.~1(b), $\alpha10$ for the frequency of 76.9~GHz is
located above 9~T, we have clearly observed part of a shoulder of
$\alpha10$ in the spectra at this frequency), as shown in Fig.~2
(top panel). We have depicted the positions of all three $\alpha9$
resonances in Fig.~1(b) with solid circles. It is tempting to
attribute these $\alpha$-resonances to the onset of the $\beta9$
transitions, which could occur if a small minority of molecules
have their easy axes tilted with respect to the majority of
molecules, whose easy axes are exactly perpendicular to applied
magnetic field, as has recently been found for Mn$_{12}$-Ac
\cite{HillCM}. However, a careful angle-dependent study of the EPR
spectra shows that the positions of $\alpha9$ and $\beta9$ exhibit
completely different angle dependences, as shown in the inset of
Fig.~2. The temperature dependence of the intensities of the
$\alpha9$ and $\beta9$ peaks reveals even more discrepancies. We
have calculated the areas under the $\alpha9$ and $\beta9$ peaks
and plotted them as a function of temperature, as represented in
Fig.~3 by solid and open circles, respectively. Again, the nature
of the $\alpha9$ resonances is different from the $\beta9$
resonances and, thus, the $\alpha9$-resonances cannot be explained
within the framework of the $S=10$ picture.

Further examination of Figs.~2 and 3 reveals that the
$\alpha9$-resonances diminish in intensity as T~$\rightarrow15$~K,
becoming invisible already at 10~K. This fact proves beyond any
doubt that the $\alpha9$ resonance originates from an excited
state of the Mn$_{12}$-BrAc molecule. We therefore conclude that,
at low frequencies, the $\alpha9$-resonance corresponds to a
transition within an excited state of Mn$_{12}$-BrAc. Within the 8
spin model described above
\cite{SessoliJACS93,Katsnelson99,YamamotoRPL02,RegnaultPRB02}, one
could imagine a low energy excitation involving the flipping of a
spin$-\frac{1}{2}$ dimer, leading to a reversal of its moment, and
to an overall $S = 9$ state for the molecule. For comparison,
Fig.~1(b) includes dotted curves corresponding to an $S = 9$
state, which were generated using precisely the same simulation as
performed for the $S = 10$ fit. For an odd total spin state, the
low field limiting behavior of odd-to-even $m$ and even-to-odd $m$
transitions is the reverse of that for an even total spin state.
Consequently, one {\em does} expect the frequency of the M$_S =
-9$ to $-8$ transition to go to zero in the low field limit within
the $S = 9$ manifold. The low frequency $\alpha9$-resonance data
lie perfectly on the $S = 9$ curves. The fit to the data for $S=9$
yields the Hamiltonian parameter $D=-0.430$~cm$^{-1}$ ($-0.62$~K),
which is thus 5\% smaller than for $S=10$. Therefore, the
anisotropy barrier for $S=9$ state is $|D|S^2 \approx 50$~K, which
is 23\% smaller than that for $S=10$ (65~K). Having established
that the low-frequency $\alpha9$-transition corresponds to an
$S=9$ state, we can estimate its approximate location relative to
$S=10$. Using the CF parameters for both $S=10$ and $S=9$ states
we were able to calculate energy levels $E_{10}(m_S)$ and
$E_9(m_S)$ for the two ground states, and both partition functions
$ Z_{10}(m_s)=\sum_{S = - 10}^{10} {e^{ - E_{10} \left( {m_S }
\right)/T} } $ and $ Z_{9}(m_s)=\sum_{S = - 9}^{9} {e^{\left( { -
E_{9} \left( {m_S } \right)+\Delta} \right)/T} } $ \noindent, for
a given temperature $T$, where $\Delta$ is the energy difference
between the bottoms of the $S=10$ and $S=9$ manifolds. The area
under the $\alpha9$ peak, at a given temperature, is proportional
to the difference of populations of the corresponding levels:

\[ A_{\alpha 9} (\Delta ,T) \propto \frac{{N_{ - 9}  - N_{ - 8}
}}{Z} = \frac{{e^{\{E_9 ( - 8) - E_9 ( - 9) \}/T} }} {{Z_{10} (T)
+ Z_9 (\Delta ,T)}} \]

\noindent Thus by varying the only parameter $\Delta$, we have
found that the $S=9$ manifold is located at $\Delta=40\pm 2$~K
above the bottom of the $S=10$ state (see inset of Fig.~3). This
implies that the supposed $S = 9$ state lies very close to the
$M_S = \pm 6$ excited state within the $S = 10$ multiplet. We have
also performed similar calculations of the temperature dependence
of the $\alpha9$-peak area for the values of $\Delta=36$~K and
$\Delta=44$~K, and both dependencies were drastically inconsistent
with our experimental data, as depicted in Fig.~3. The obtained
location of the $S=9$ excited state at $\Delta =40\pm2$~K is in
perfect agreement with recent calculations~\cite{Park03}.

In summary, detailed frequency and temperature dependent EPR
studies of Mn$_{12}$-Ac reveal the existence of an $S=9$ state
located only $40\pm2$~K above the $S = 10$, M$_S = \pm 10$ ground
state. This result is in perfect agreement with theoretical
predictions~\cite{Katsnelson99,Park03}. The effects of the
co-existence of an excited $S=9$ state and the ground $S=10$ state
in the Mn$_{12}$ molecule are not known, and we hope these
investigations will stimulate further theoretical studies. Our
studies also indicate that Mn$_{12}$-BrAc is an intrinsically
cleaner system than Mn$_{12}$-Ac, which we believe to be connected
with the fact that the former possesses a full compliment of four
solvent molecules per formula unit. Thus, future investigations of
the title compound may provide further insights into the quantum
magnetization dynamics of giant spin ($S=10$) SMMs.

We thank N.~S.~Dalal, E.~Rumberger, D.~N.~Hendrickson,
E.~del~Barco, and A.~D.~Kent for useful discussion. This work was
supported by the NSF (DMR0103290, DMR0239481, and CHE0123603).
S.~H. acknowledges the Research Corporation for financial support.

\clearpage

\noindent{{\bf Figure captions}}

\bigskip

FIG.~1. (a) Angle dependence of the EPR spectra of Mn$_{12}$-BrAc
in the range of $\pm 3.6^{\circ}$ either side of the hard plane,
with an angular step of $0.36^{\circ}$. (b) The SIM~\cite{Weihe}
simulations of the EPR spectra for the field tilted up to $\pm
3.6^\circ$ away from the hard plane. (c) Fits to Eq.~1 for the
frequency dependence of the hard plane spectra for the $S=10$
state (solid curves, open circles) and for the $S=9$ state (dotted
curves, solid circles); the CF parameters for this simulation are
given in the text. Open and closed circles are experimental data
at frequencies of 51.5, 65.4, and 76.9~GHz.

\bigskip

FIG.~2. (Top panel) Typical single-crystal EPR spectra of
Mn$_{12}$-BrAc at 65.4 and 76.9 GHz ($\mathbf{B} \perp z$); the
$\alpha9$ resonance is evidenced at high fields for both
frequencies. (Bottom panel) Temperature dependence of the EPR
spectra of Mn$_{12}$-BrAc at 51.5~GHz ($\mathbf{B} \perp z$).
Inset: angle dependence of several of the most important
resonances.

\bigskip

FIG.~3. Temperature dependence of the area of the $\alpha9$ (solid
circles) and $\beta9$ (open circles) resonances. The solid line
through the open circles is a guide to eye. The curves through the
solid circles represent the calculated $\alpha9$-resonance areas
assuming $\Delta=36$~K (dashed), $\Delta=40$~K (solid), and
$\Delta=44$~K (dotted). Inset: schematic for the energy levels of
both the $S=10$ and $S=9$ states in zero magnetic field. $S=9$ is
located at an energy $\Delta=40\pm 2$~K above the bottom of the
$S=10$ state.

\clearpage
\begin{figure}
\includegraphics{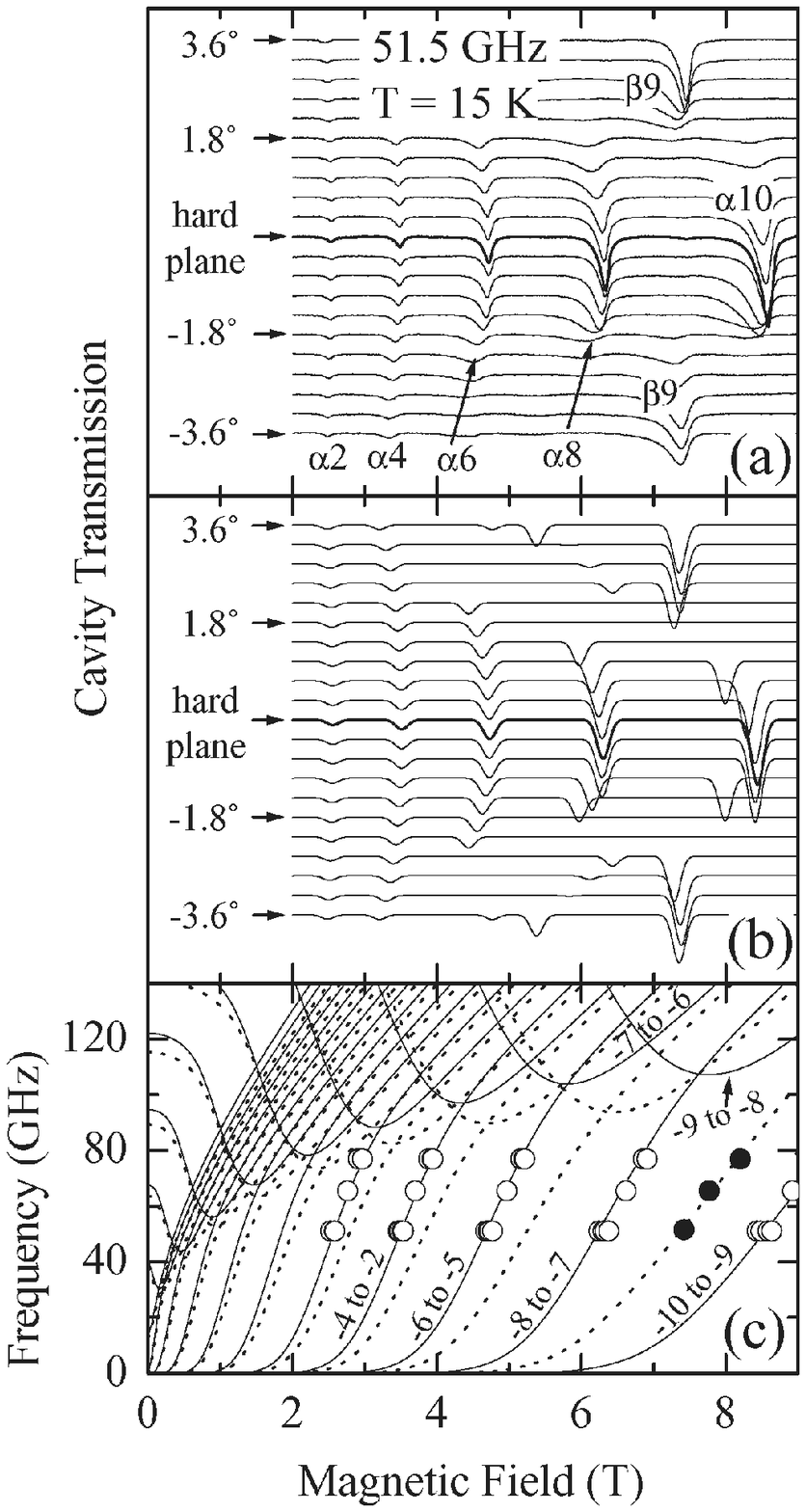}
\caption{\label{fig1} K.~Petukhov {\em et al.}}
\end{figure}

\bigskip
\begin{figure}
\includegraphics{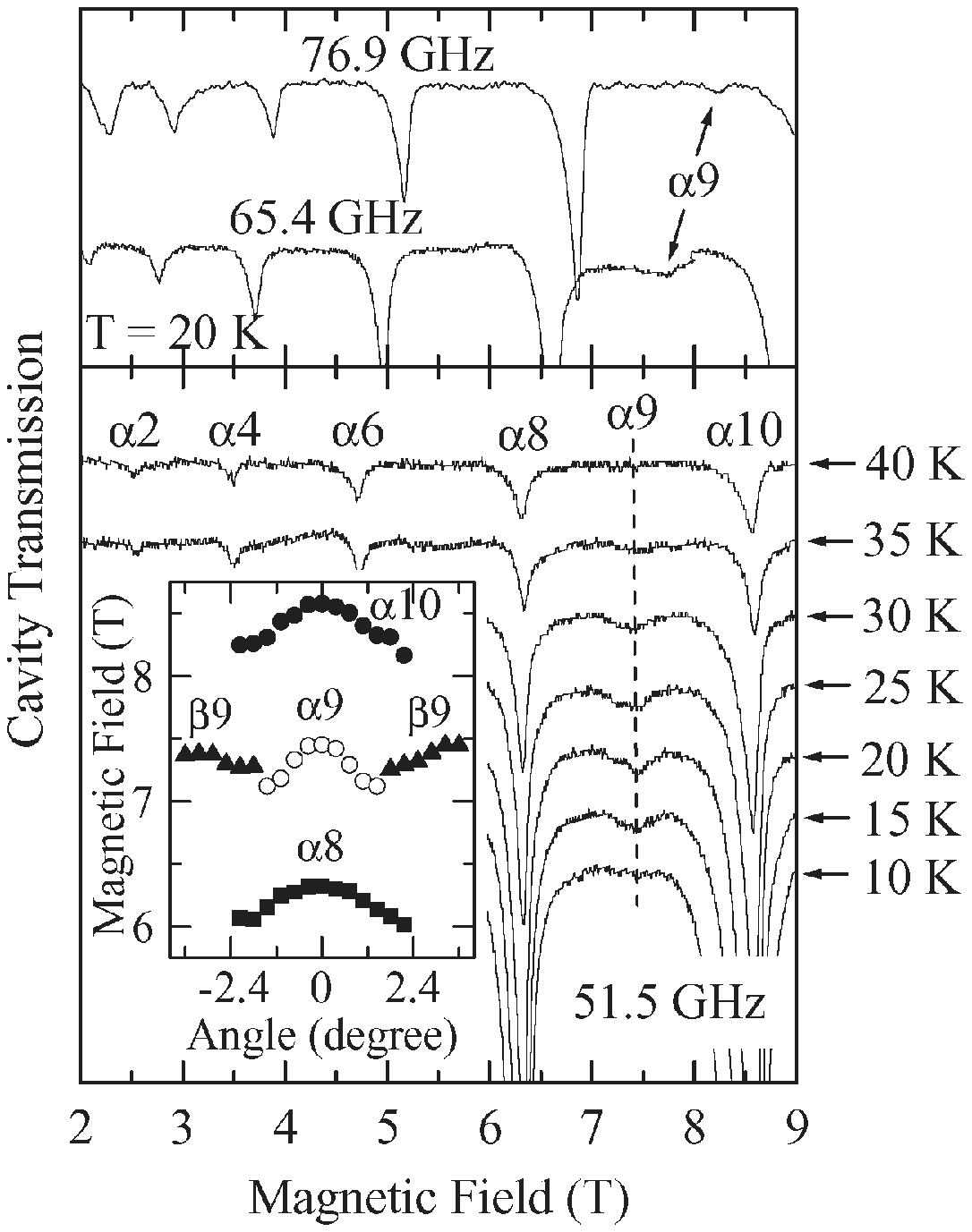}
\caption{\label{fig2} K.~Petukhov {\em et al.}}
\end{figure}

\bigskip
\begin{figure}
\includegraphics{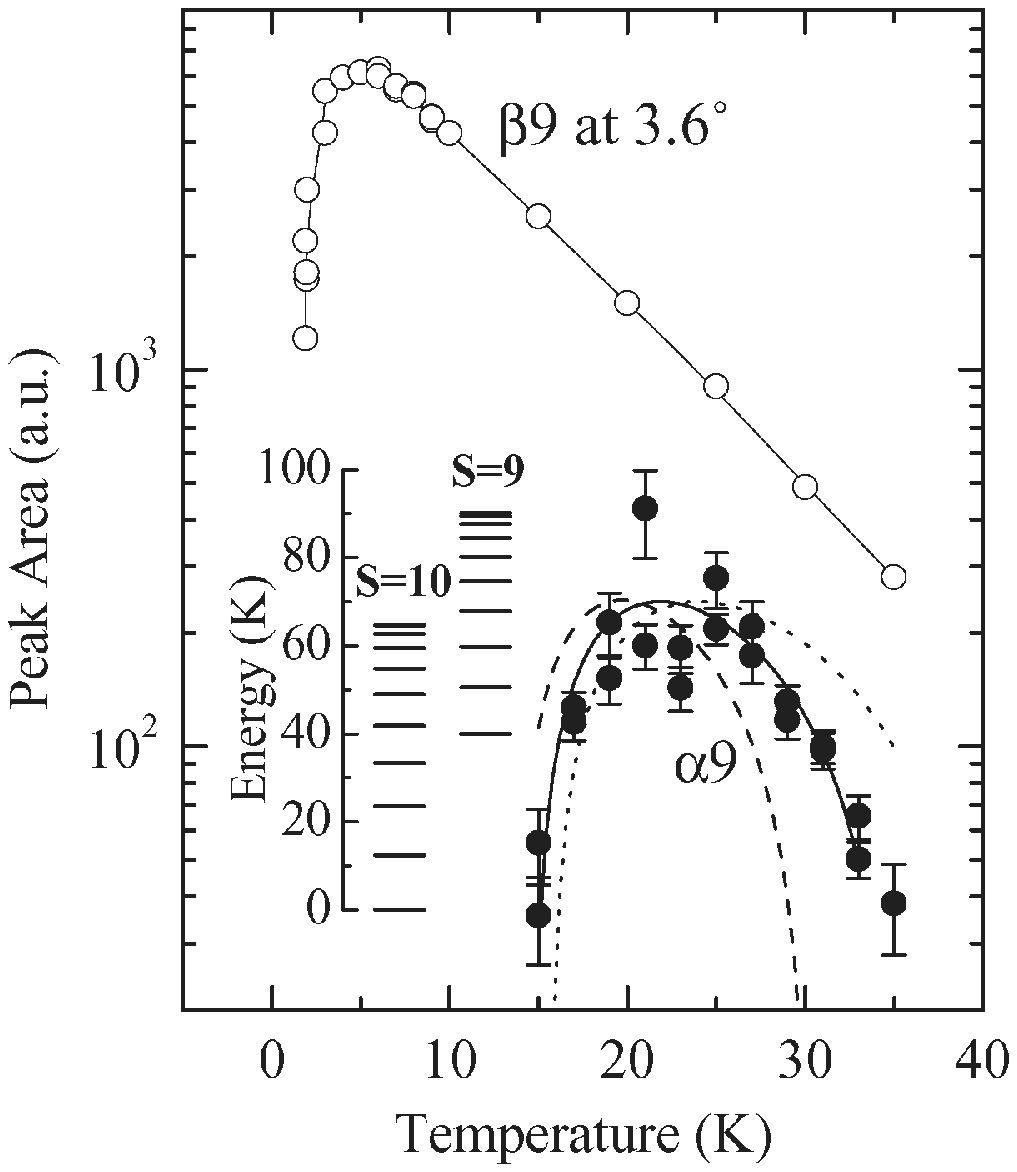}
\caption{\label{fig3} K.~Petukhov {\em et al.}}
\end{figure}


\begin{thebibliography}{39}
\bibitem{note1}
We note that this numbering convention differs slightly from the
scheme used in earlier publications.~We previously used all
integers when numbering the $\alpha$ and $\beta$
resonances~\cite{HillPRL98}.~This scheme ignores the obvious
symmetries associated with the various transitions.~Therefore, we
now use only even integers for $\alpha$ resonances, and odd
integers for $\beta$ resonances, i.e.~$\alpha9$ under the old
notation is now $\alpha8$, $\beta10$ is $\beta9$, $\alpha8$ is
$\alpha6$, and so on
\end{thebibliography}
\end{document}